\documentclass[preprints,preprints,article,accept,moreauthors,pdftex]{Definitions/mdpi} 
\usepackage{bm}
\usepackage{listings}
\usepackage{subfigure}
\usepackage{amsmath}
\usepackage{amssymb}
\usepackage{booktabs}
\firstpage{1} 
\pubvolume{xx}
\issuenum{1}
\articlenumber{1}
\pubyear{2019}
\copyrightyear{2019}
\history{Received: December 2019}

\newcommand{\floor}{\mathop{\mathrm{floor}}}

\Title{Equivalence between forward rate interpolations and discount factor interpolations for the yield curve construction}
\Author{Jherek Healy}
\AuthorNames{Jherek Healy}
\address{}
\corres{Correspondence: jherekhealy@protonmail.com}
\abstract{The traditional way of building a yield curve is to choose an interpolation on discount factors, implied by the market tradable instruments. Since then, constructions based on specific interpolations of the forward rates have become the trend. We show here that some popular interpolation methods on the forward rates correspond exactly to classical interpolation methods on discount factors. This paper also aims at clarifying the differences between interpolations in terms of discount factors, instantaneous forward rates, discrete forward rates, and constant period forward rates. }

\begin{document}

\section{Introduction}
The traditional way of building a yield curve is to choose an interpolation on discount factors, implied by the market tradable instruments \citep{ametrano2009bootstrapping,andersen2010interest}. Since then, with the introduction of the monotone convex interpolation by \citet{hagan2006interpolation}, constructions based on specific interpolations of the discrete forward rates have become the trend, mostly because the discrete forward rates are more directly related to the market observables. We show here that some popular interpolation methods on the consecutive discrete forward rates, or on instantaneous forward rates, correspond actually to  classic interpolation methods on discount factors.

The equivalence was suggested in \citep{lefloch2013stable}. The aim of this paper is to clarify this equivalence further, and to show in particular that the smart quadratic interpolation, often used in banks, and described in \citep{hagan2018building} is equivalent to a Hermite spline interpolation on the logarithm of discount factors, and the area-preserving quadratic spline interpolation of \citet{hagan2018building} is equivalent to a cubic spline interpolation of class $C^2$ on the logarithm of discount factors.

Since the 2008 financial crisis, there is not a single curve used for discounting or projecting rates anymore, but multiple, jointly calibrated, curves. Distinct curves are used for discounting or for projecting rates of specific tenors and currency. In this context, cubic spline interpolations may be applied directly to discrete forward rates of constant period as in \citep{henrard2014interest}. We show that additional knots must be introduced to derive the equivalent cubic spline interpolation in terms of the logarithm of pseudo-discount factors.

To conclude, we present a few alternative piecewise cubic interpolation schemes with natural shape preserving properties.

\section{Yield curve construction}\label{sec:yield_curve_construction}
We start by introducing the concepts of discount factors, discrete forward rate and instantaneous forward rate, key for the curve construction.
Let $P(t)$ denote the price of a zero coupon bond delivering for certain \$1 at maturity $t$. The continuously compounded yield $y$ is given by
\begin{equation}P(t)=e^{-y(t)t}\,.\end{equation}
The instantaneous forward rate $f$ is defined by
\begin{equation}
P(t) = e^{-\int_0^t f(u) du}\,.\label{eqn:inst_forward_rate}
\end{equation}
The forward rate $f_i^d$ from $t_{i-1}$ to $t_i$, which corresponds to the rate fixed at time 0 to borrow from $t_{i-1}$ and $t_{i}$ is defined by 
\begin{equation*}
P(t_{i-1}) e^{-f_i^d (t_{i}-t_{i-1})} = P(t_i) \,,
\end{equation*}
or equivalently,
\begin{equation}
f_i^d= -\frac{\ln P(t_i)- \ln P(t_{i-1}) }{t_{i}-t_{i-1}}\,.\label{eqn:discrete_forward_rate}
\end{equation}

From equations \ref{eqn:inst_forward_rate} and \ref{eqn:discrete_forward_rate}, we deduce that discrete forward rate corresponds to the area of the instantaneous forward rate between $t_{i-1}$ and $t_i$. We have
\begin{equation}
f_i^d = \frac{1}{t_{i}-t_{i-1}}\int_{t_{i-1}}^{t_{i}} f(u)du\,.\label{eqn:discrete_forward_rate_integral}
\end{equation}

Let $z$ be the logarithm of the discount factors:
\begin{equation}z(t)= \ln P(t) = -y(t)t\,.\end{equation}
The relation between the logarithm of the discount factor and the instantaneous forward rate is then
\begin{equation}
f(t)=-\frac{\partial \ln P }{\partial t} (t) = -\frac{\partial z}{\partial t} (t)\,,
\end{equation}
and the relationship with the discrete forward rate is
\begin{equation}
f_i^d= -\frac{z(t_i)- z(t_{i-1}) }{t_{i}-t_{i-1}}\,.\label{eqn:discrete_forward_rate_z}
\end{equation}

In the above equations, we followed \citet{hagan2006interpolation} and defined $f_i^d$ as a continuously compounded rate. Yet, it still represents effectively a discrete forward rate between two dates, in contrast with the instantaneous forward rate defined by Equation (\ref{eqn:inst_forward_rate}). In particular, the continuously compounded rate $f_i^d$ is equivalent to, and may be trivially converted into a specific single period discrete rate.  We will discuss the latter in more details, in Section \ref{sec:henrard_interp}.

The yield curve is the curve followed by $y(t)$ at any time $t$, or equivalently the curve which defines the (pseudo-)zero coupon bond value $P(t)$. The market quotes a limited number of securities, whose prices depend on the (pseudo-)zero coupon bond values at a discrete set of dates $(t_j)_{j=1,...,M}$. A proper yield curve must be able to reprice the market securities. A complete description of yield curve construction is given in \citep{ametrano2009bootstrapping, andersen2010interest}. Let $V_i$ be the price of $N$ securities. Typically, for a Libor curve, those securities are Libor deposits for the first few months, Eurodollar futures for up to 3 or 4 years and par swaps for the rest of the curve.
We assume that the securities can be written as a linear combination of discount bond prices:
\begin{equation}
V_i = \sum\limits_{j=1}^M c_{i,j} P(t_j) \texttt{ , } i=1,...,N \label{eqn:securities}
\end{equation}
with $t_1,...,t_M$ a finite set of dates, in practice corresponding to the cash flow dates of the $N$ benchmark securities. In the context of multiple curves construction, the coefficients $c_{i,j}$ may depend on pseudo-discount factors obtained from the other curves. In the case of the OIS curve, the price of some securities is not necessarily a linear combination of a discrete set of discount factors, but only a non-linear function of this set. We will detail this in Section \ref{sec:ois_curve}.

If we want to calibrate a set of discount factors to the market instruments, we only need to solve a linear system in the discount factors $(P(t_j))_{j=1,...M}$. In reality, the interpolation plays however a role, since the number of securities $N$ is typically smaller than the number of discount factors $M$ and the system is under-determined. The interpolation allows then to restrict the possible shapes allowed to a space of shapes implied by $N$ parameters. The typical yield curve calibration algorithm consists in solving for the $N$ parameters of a given interpolation function so that Equation (\ref{eqn:securities}) holds. This is not a linear problem anymore, as the interpolation function is not a linear in the discount factors and a non-linear solver (Levenberg-Marquardt in our numerical examples) must be used. The linearity property of Equation (\ref{eqn:securities}) is effectively not used in the calibration.

\section{Hagan and West interpolation}
Let us recall Hagan and West monotone convex spline construction \citep{hagan2006interpolation}. 

Firstly, a suitable set of forward rates is computed according to the following procedure.
Let $f_i^d$ be the input discrete forward rate at node $i$, the rate at point $t_i$ is defined for $i=1, 2, ..., n-1$ by:
\begin{align}
f_i &= f(t_i) = \frac{t_i-t_{i-1}}{t_{i+1}-t_{i-1}} f_{i+1}^d+ \frac{t_{i+1}-t_i}{t_{i+1}-t_{i-1}} f_i^d\,, \label{haganwest_forward}\\
f_0 &=f(t_0)= f_1^d-\frac{1}{2} (f_1-f_1^d)\,, \label{haganwest_forward0}\\
f_n &= f(t_n)=f_n^d-\frac{1}{2} (f_{n-1}-f_n^d)\,. \label{haganwest_forwardn}
\end{align}
Secondly, for $i=1,2,...,n$, let $g_i(x)=f\left(t_{i-1}+(t_i-t_{i-1})x\right)-f_i^d$. The quadratic $g$ is defined for $x \in [0,1]$ by
\begin{align*}
g_i(x) &= g_i(0) (1-4x+3x^2)+g_i(1)(-2x+3x^2)\,.
\end{align*}
The instantaneous forward rate is computed from $g_i$ through $f(t)=g_i\left(\frac{t-t_{i-1}}{t_i-t_{i-1}}\right) + f_i^d$, for $t \in [t_{i-1},t_i)$.
This definition creates a continuous interpolation which preserves the consecutive discrete forward rates as we have $\frac{1}{t_i-t_{i-1}} \int_{t_{i-1}}^{t_i}f(t)dt = f_i^d$ by construction. 

Secondly, the function $g$ is modified so that the  positivity and monotonicity of the instantaneous forward rates is preserved. Since the advent of negative rates, and because those modifications may create unstable hedges in some circumstances \citep{lefloch2013stable}, practitioners often discard the modifications to $g$. A recent paper by \citet{hagan2018building} also presents this same interpolation (without the modifications for monotonicity and positivity) with the name "smart quadratic" and suggests it is close to what Bloomberg was using internally at the time. 

\section{Cubic Hermite spline interpolation}
Given the data $z(t_0), z(t_1), ..., z(t_n)$ with $t_0 < t_1 < ... < t_n$, following \citep[p. 39--32]{de1978practical}, 
a  piecewise cubic interpolant $p$ is of the following form 
for $i \in 0,...,n-1$, for  $t \in [t_i,t_{i+1}]$,
\begin{align}
p(t) &= p_i(t) = c_{i,0}+c_{i,1}(t-t_i)+c_{i,2}(t-t_i)^2+c_{i,3}(t-t_i)^3\,,\label{eqn:hermite_spline}
\end{align}
with coefficients $c_{i,j} \in \mathbb{R}$.
The conditions for an interpolation of class $C^1$ are:
\begin{align}\label{hermite_spline}
p_i(t_i) &= z(t_i),\;\;\; p_i'(t_i) = s_i,\;\;\; i = 0,...,n \\
p_i(t_{i+1}) &= z(t_{i+1}),\;\;\; p_i'(t_{i+1}) = s_{i+1},\;\;\; i = 0,...,n-1
\end{align}
where $s_{i}$ are free parameters.
Let
\begin{align}
d_{i} &= \frac{z(t_{i+1})-z(t_i)}{t_{i+1}-t_{i}}\label{bessel_S}\,.
\end{align} 
The interpolation conditions give:
\begin{align}
c_{i,0}(t_i) &= z(t_i)\,,\\
c_{i,1} &= s_i\,,\\
c_{i_2} &= \frac{3d_{i}-s_{i+1}-2s_i}{t_{i+1}-t_i}\,,\label{eqn:hermitec2}\\
c_{i_3} &= -\frac{2d_{i}-s_{i+1}-s_i}{(t_{i+1}-t_i)^2}\,.\label{eqn:hermitec3}
\end{align}

For local interpolation schemes, the $s_i$ are chosen so that the $i$th cubic polynomial depends only on information from, or near the interval $[t_i,t_{i+1}]$. While the $s_i$ directly correspond to the slope of the interpolation at $t_i$, the variables $d_i$ are just used as a convenient notation.

Now if we let $z(t_i)$ correspond to the logarithm of the discount factor at date $t_i$, per Equation (\ref{eqn:discrete_forward_rate_z}), we have then the identity \begin{equation}d_i=-f_{i+1}^d\,.\end{equation}

In a Bessel spline, for $i=1,...,n-1$, the $s_i$ are chosen to be the slope of the parabola interpolating three consecutive data-points. Its order of accuracy is $O(\delta t^2)$.
\begin{align}
s_i &= \frac{(t_i-t_{i-1})d_i+(t_{i+1}-t_i)d_{i-1}}{t_{i+1}-t_{i-1}}\label{bessel_s} \,.
\end{align}
The remaining $s_0$ and $s_n$ are determined by an appropriate choice of boundary conditions. 
From equations \ref{bessel_s} and \ref{haganwest_forward}, along with the identity $d_i=-f_{i+1}^d$, we obtain the identity 
\begin{equation}
s_i = -f_i  \quad \textmd{for } i=1,...,n-1\,.
\end{equation}

A typical boundary condition is the so-called natural boundary condition where $p''(t_0) = 0 = p''(t_n)$. According to Equation (\ref{eqn:hermitec2}), for the left boundary $t_0$, this corresponds to 
\begin{equation*}
\frac{3d_{0}-s_{1}-2s_0}{t_{1}-t_0}  = 0\,,
\end{equation*}
or equivalently \begin{equation}
s_0 = d_0 - \frac{1}{2}(s_1 -d_0)\,.
\end{equation}
Similarly for the right boundary $t_n$, according to equations \ref{eqn:hermitec2} and \ref{eqn:hermitec3},we have
\begin{align}
2\frac{3d_{n-1}-s_{n}-2s_{n-1}}{t_{n}-t_{n-1}} - 6 \frac{2d_{n-1}-s_{n}-s_{n-1}}{t_{n}-t_{n-1}} &= 0\,,
\end{align}
or equivalently 
\begin{equation} 
s_n = d_{n-1} - \frac{1}{2}(s_{n-1} -d_{n-1})\,.
\end{equation}
The boundaries thus corresponds to the boundaries set in the forward rate interpolation defined by equations \ref{haganwest_forward0} and \ref{haganwest_forwardn} and we thus also have $s_0 = -f_0$ and $s_n=-f_n$.

The derivative of the cubic Hermite spline $-p$ is a continuous quadratic spline, which interpolates $f_i$ at $t_i$ for $i=0,...,n$. It must thus be the same quadratic spline as Hagan and West. This may also be verified by an explicit calculation of $p'$.

The cubic Hermite spline on the logarithm of discount factors will also preserve the area from $t_i$ to $t_{i+1}$ since, by construction, we impose $p(t_i)=z(t_i)$ and we have( $z(t_{i})-z(t_{i-1}) = -\int_{t_{i-1}}^{t_i} f(u) du$ according to equations \ref{eqn:discrete_forward_rate_integral} and \ref{eqn:discrete_forward_rate_z}.

The flat extrapolation of the forward rates for $t < t_0$ or $t_n < t$ corresponds to a $C^1$ linear extrapolation in the logarithm of discount factors.

\section{Hagan smoother area preserving interpolation}\label{sec:hagan_smoother}
In the area preserving quadratic spline interpolation of \citet{hagan2018building}, the following spline is considered for the instantaneous forward:
\begin{equation}
f(t) = f_{i-1} (1-x_i(t)) + f_i x_i(t) - 3 (f_{i-1}+f_i+2 f_i^d) x_i(t) (1-x_i(t)) \quad\textmd{ for } t \in [t_{i-1},t_i)\label{eqn:hagan_api_spline}
\end{equation}
with $x_i(t) = \frac{t-t_{i-1}}{t_i-t_{i-1}}$.
The $f_i$ are not defined by Equations (\ref{haganwest_forward}) anymore but are chosen so that $f'(t_i^-) = f'(t_i^+)$ and the boundaries are defined by $f'(t_0)=0$ and $f'(t_n)=0$. This leads to a tridiagonal system on the $f_i$.

Let us now explore its equivalence to a cubic spline on the logarithms of discount factors.
The first derivative of the Hermite cubic spline defined by Equation (\ref{eqn:hermite_spline}) is
\begin{equation}
p'(t) = c_{i-1,1} + 2 c_{i-1,2} (t-t_{i-1}) + 3 c_{i-1,3} (t-t_{i-1})^2 \quad\textmd{ for } t \in [t_{i-1},t_i)
\end{equation}
with $c_{i,1} = s_i$ and $c_{i,2}, c_{i,3}$ defined by equations \ref{eqn:hermitec2}, \ref{eqn:hermitec3}.

 When the Hermite cubic spline is applied to the logarithm of discount factors $z(t_i)$, we have $s_i = -f_i$ and by rewriting the Equation (\ref{eqn:hagan_api_spline}) in the same form, it can easily be verified that $p'(t) = -f(t)$.
The condition $f'(t_i^-) = f'(t_i^+)$ correspond to $p''(t_i^-) = p''(t_i^+)$ and the boundary conditions correspond to the natural boundary conditions.
Such a spline is thus the standard cubic spline of class $C^2$ as described in \citep[p. 43]{de1978practical}. The derivative of the cubic spline $-p$ is a $C^1$ quadratic spline, which interpolates $f_i$ at $t_i$ for $i=0,...,n$. It must thus be the same area preserving quadratic spline as Hagan.

It is well known, that on some monotone or convex input data, a standard cubic spline interpolation may oscillate \citep{dougherty1989nonnegativity,huynh1993accurate}. A typical example is data from a Heaviside like function. If the input data were directly zero rates, this would happen on many real world examples. In our case, the input data are the logarithm of discount factors, that is, zero rates multiplied by their corresponding maturity. This multiplication will, in effect, considerably smooth out the input data. In order to reproduce a clear oscillation, much larger zero rates variations are needed, so large, that they are unlikely to be realistic at all.


\section{Cubic spline interpolation on the discrete forward rates}\label{sec:henrard_interp}
In a multi-curve environment, each curve is associated to a specific index, and a specific tenor.
For a given index and tenor, instead of interpolating the logarithm of discount factors, or the instantaneous forward rates, \citet{henrard2014interest} proposes to interpolate the discrete forward rates with start dates $u$ and maturity dates $v$ defined in terms of pseudo-discount factors $P$ by
\begin{equation}
\hat{f}^d(u,v)  = \frac{1}{\delta}\left(\frac{P(u)}{P(v)}-1\right) \label{eqn:discrete_fwd_one}
\end{equation}
so that the price of an Ibor coupon with start date $u$ and maturity $v$  is $P_D(v)\delta \hat{f}^d(u,v)$, with $P_D$ being the discount factor associated to the relevant discount curve, and $\delta$ the accrual factor for the period. In a multi-curve environment, $P_D$ is different from $P$.

In order to be consistent with our previous notation, we transform the one-period discrete rate $\hat{f}^d$ into a continuously compounded rate $f^d$ through the relation
\begin{equation}
e^{f^d(u,v) (v-u)} = \hat{f}^d(u,v) \delta + 1\,.
\end{equation}
In terms of $f^d$, Equation (\ref{eqn:discrete_fwd_one}) becomes 
\begin{equation}
f^d (u,v) =  - \frac{\ln P(v) - \ln P(u)}{v-u}\,. \label{eqn:discrete_fwd_cnt}
\end{equation}
In particular, we have $f^d(t_{i-1},t_i)= f_i^d$ where $f_i^d$ is defined by Equation (\ref{eqn:discrete_forward_rate_integral}).

\subsection{From discount factors to discrete forward rates}
Let $p$ be a $C^2$ cubic spline interpolation on the logarithm of pseudo-discount factors at the dates $t_i$ as in Section \ref{sec:hagan_smoother}. From Equation \ref{eqn:discrete_forward_rate_integral}, we have
\begin{equation}
f^d (t, t+\Delta)  = \frac{ p(t+\Delta)-p(t)}{\Delta}\,.\label{eqn:fd_p}
\end{equation}
In the discrete forward curve construction, the discrete forward rates used as input of the interpolation are all of the same tenor, and thus $\Delta$ is kept constant\footnote{We neglect any mismatch related to week-ends or holidays.} for all $t$. A priori, the set of observations times $t_i$ is however not evenly distributed.

If we start from pseudo-discount factors observation times $(t_i)$ and want to construct an equivalent discrete forward interpolation, we also need the pseudo-discount factors at times $(t_i + \Delta)$. Let $(\tau_j)_{j=1,...m}$ be the sorted set of dates such that $(\tau_j)_{j=1,...m} = (t_i)_{i=1,...,n} \cup ({t_i +\Delta})_{i=1,...,n}$ and let us define the interpolation function $\bar{f}^d$ by \begin{equation}\bar{f}^d(t) = f(t, t+\Delta)\,.\label{eqn:bf_definition}\end{equation} 
We know that $p$ is a $C^2$ cubic piecewise polynomial with knots $(t_i)_{i=1,...,n}$, and thus $\bar{f}^d$ is also a $C^2$ cubic piecewise polynomial with knots $(\tau_j)_{j=1,...m}$. Furthermore, we have by construction
\begin{equation}
\bar{f}^d(t_i)=\frac{ p(t_i+\Delta)-p(t_i)}{\Delta}\,.\label{eqn:f_p_1}
\end{equation}
We have thus shown that a $C^2$ cubic spline interpolation on the logarithm of the pseudo-discount factors at knots $(t_i)_{i=1,...,n}$ implies the same discrete forward rates at all times $t$ as a $C^2$ cubic spline interpolation on the continuously compounded discrete forward rates at knots $(\tau_j)_{j=1,...m}$, assuming a constant period length $\Delta$ for the continuous compounding. The equivalent interpolation in terms of discrete forward rates introduces additional discontinuities in the third derivative of $\bar{f}^d$, located at the knots $(t_i + \Delta)$. 

On the interval $(t_i, t_{i+1}-\Delta)$, the same piecewise polynomial $p_i$ will be used to compute the difference and Equation (\ref{eqn:fd_p}) simplifies to
\begin{equation}
\bar{f}^d(t)= p_i'(t) + \frac{\Delta}{2} p_i''(t) + \frac{\Delta^2}{6} p_i'''(t)\,.\label{eqn:fd_quad}
\end{equation}
If $p_i$ is cubic in $t$, the discrete forward rate will thus be quadratic in $t$. In contrast, on the interval $(t_{i+1}-\Delta,t_{i+1})$, two different piecewise polynomials will be used, and the discrete forward rate will, a priori, not be quadratic, but stay cubic.

The choice of interpolating on pseudo-discount factors with a spline thus implies a less smooth spline interpolation in terms of discrete forward rates when compared to a direct spline interpolation on the discrete forward rates. Hence a direct interpolation in terms of discrete forward rates may be preferable, if all we need to compute the yield curve instrument prices are only discrete forward rates.

As evidenced in \citep{clarus2016adapting}, there is however an additional subtlety when interpolating directly on the discrete forward rates, related to the ambiguity of the start date, in relation with a given end date. Indeed, because of holidays and week-ends, in a standard modified following business day convention, a given end date (for example a Monday) will correspond to multiple start dates, assuming a constant period for all the forward rates (for example three months). Using the end date as variable to interpolate on would create an ambiguity. A simple way to resolve this issue is to define the interpolation on the start dates instead as in  Equation (\ref{eqn:bf_definition}) .

\subsection{From discrete forward rates to discount factors}
If we start from discrete forward observation dates $(t_i)$, and assume a $C^2$ cubic piecewise polynomial representation on the knots $(t_i)$ for $\bar{f}^d$, the values at $t_i - k \Delta$ for $0 \leq k \leq \floor\left(\frac{t_i}{\Delta}\right)$ are also needed to establish the equivalent $C^2$ cubic piecewise interpolation in terms of pseudo discount factors, as, by summing Equation (\ref{eqn:f_p_1}), we obtain the relation
\begin{equation}
p(t) - p(t-K\Delta)= {\Delta}\sum_{k=0}^{K}\bar{f}^d(t - k \Delta)\,,
\end{equation}
with $K=\floor\left(\frac{t}{\Delta}\right)$. The value at $p(t-K\Delta)$ may be determined by the choice of a specific extrapolation of $\bar{f}^d$ for $t<\Delta$. For example, for a constant extrapolation in $\bar{f}^d$, $p(t-K\Delta) = (t-K\Delta) f^d(\Delta)$ with the convention $p(0) = 0$ (the discount factor at time zero is one). The function $p$ will thus be a $C^2$ piecewise polynomial on $(\tau_j)_{j=1,...,m} =\cup_{i=1,...,n} \cup_{0 \leq k \leq \floor\left(\frac{t_i}{\Delta}\right)} (t_i - k \Delta)$. Discontinuities in the third derivative of $\bar{f}^d$ at $t_i$ will translate to discontinuities in $p'''$ at $t_i - k \Delta$ for $0 \leq k \leq \floor\left(\frac{t_i}{\Delta}\right)$.

\subsection{The specific case of the OIS curve}\label{sec:ois_curve}
The USD Overnight Index Swap (OIS) curve is typically built from the Fed fund spot rates, Fed fund futures, and Fed fund OIS swaps quotes (see Table \ref{tbl:fedfund_curve} for an example). All those instruments involve only the discrete one day forward rates. The OIS curve thus allows to capture this one day forward rate, across maturities. If we were to interpolate directly this discrete forward rate, the pricing of a 50y OIS swap would involve to interpolate the forward rate value every business day during 50 years, as a coupon of the OIS swap is calculated by
\begin{equation*}
C = N \prod_{j=1}^{l} (1 - r(t_j) \delta_j) - N r_{\textsf{par}} \delta\,,
\end{equation*}
where $N$ is the notional amount, $r(t)$ is the effective OIS rate for the date $t$, $\delta_j$ is the accrual period at $t_j$, which is one day (except around non-business days where the accrual period is adjusted). Finally, $r_{\textsf{par}}$ represents the interest rate of the fixed leg, $\delta$ is the coupon period (typically one year) and is defined in the Actual/360 daycount convention. We assume that a coupon period consists of $l$ accrual periods.
In terms of one day discrete forward rates $\hat{f}^d$, the corresponding price for this coupon is
\begin{equation}
C = N P(T) \left[\prod_{j=1}^{l} (1 + \hat{f}^d(t_j) \delta_j) - r_{\textsf{par}} \delta\right]\,,
\end{equation}
where $T$ is the coupon payment date. Using Equation (\ref{eqn:discrete_fwd_one}), the coupon price may be also expressed in terms of pseudo discount factors:
\begin{equation*}
C = N P(T) \left[ \prod_{j=1}^{l} \frac{P(t_j)}{P(t_{j+1})} - r_{\textsf{par}} \delta \right]\,.
\end{equation*}
If we neglect the payment lag, the above equation simplifies to
\begin{equation}
C = N \left[P(t_1) - r_{\textsf{par}} \delta P(T)\right]\,.
\end{equation}
The pricing of an OIS swap is much more efficient and practical with pseudo-discount factors, as those directly record the accrued rate. It will thus be preferable to interpolate on the pseudo-discount factors, even if the resulting interpolation is less smooth than an interpolation on the discrete one-day forward rates.

Fed fund futures do not involve the compounded rate, but the arithmetic average rate over the futures period. The price $F$ of one contract is $F = 100 - R$ with $R$ being the arithmetic average\footnote{In the settlement, $R$ is rounded to the nearest one-tenth of one basis point.} of daily effective funds rates during the contract months:
\begin{equation}
R = \frac{\sum_{j=1}^l r(t_j) \delta_j}{\sum_{j=1}^l \delta_j}\,.
\end{equation}
In terms of discount factors the average rate is
\begin{equation}
R = \frac{\sum_{j=1}^l \frac{P(t_{j})}{P(t_{j+1})}- 1 }{\sum_{j=1}^l \delta_j}\,.
\end{equation}
The equation does not simplify, and all discount factors from the start date to the end date of the future contract are needed. It is not problematic since the Fed fund future's period is one month, and thus less than 30 discount factors are needed. Note that, contrary to the usual curve instruments\footnote{This is also the case for the Fed fund basis swap, which also involves an arithmetic average.}, the price of a Fed fund future is not a linear combination of discount factors. As explained in section \ref{sec:yield_curve_construction}, the usual algorithm to calibrate the yield curve may however still be applied.

\begin{figure}
	\begin{center}  
		\subfigure[1d forward rate]{
			\includegraphics[width=\textwidth]{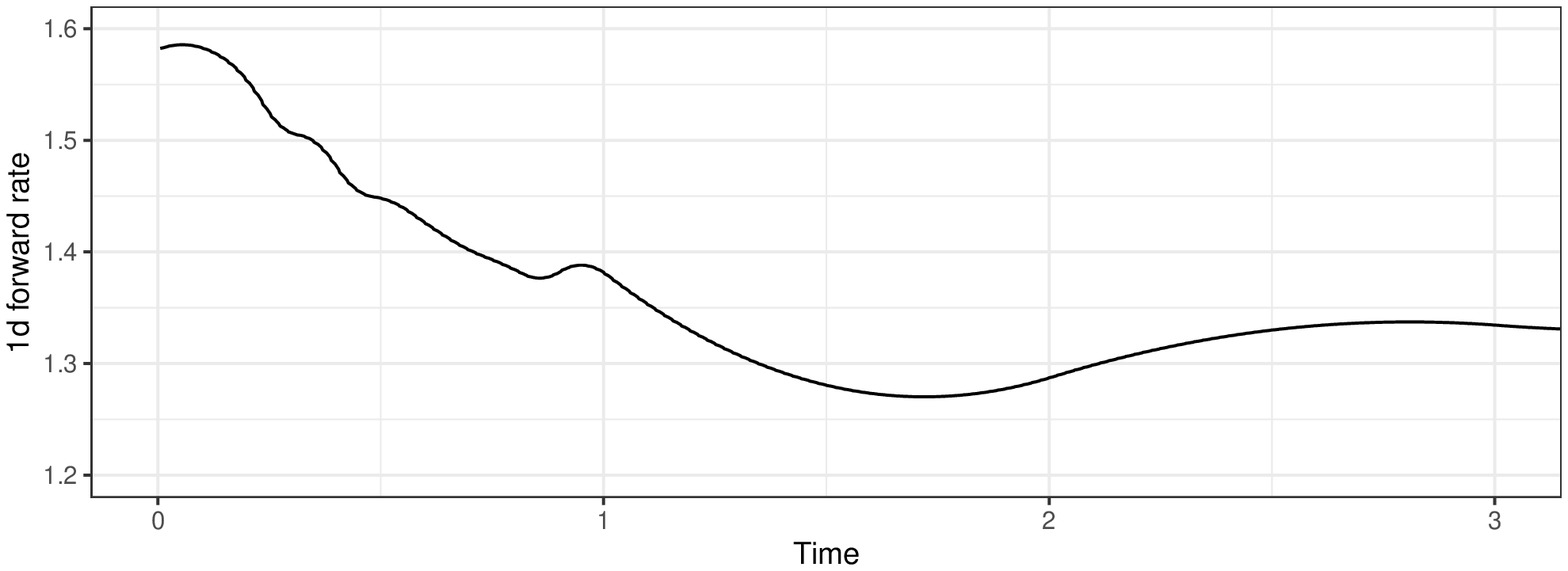}}\\
		\subfigure[Second derivative]{
			\includegraphics[width=\textwidth]{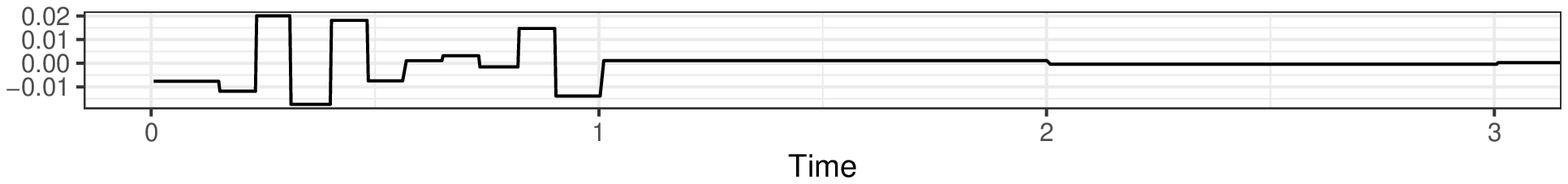}}
		\caption{\label{fig:ff_curve}Fed fund curve using the data from Table \ref{tbl:fedfund_curve}, zoomed between zero and three years.}%
	\end{center}
\end{figure}

In Figure \ref{fig:ff_curve}, we plot the one day forward rate, implied by a cubic spline interpolation of class $C^2$ on the pseudo-discount factors. The second derivative of the one day forward rate is flat, and appears to jump at the knots. The jump represents the change of interpolation in the interval $(t_i, t_i-\Delta)$ from a quadratic to a cubic (see Equation \ref{eqn:fd_quad}): the second derivative is actually continuous, but changes significantly in a short period of time $\Delta$.

\section{Alternative interpolations}
\subsection{Histosplines}
What Hagan calls an area preserving spline is also known as a \emph{histospline} in the literature \citep{morandi1989piecewise,costantini2007shape}. More generally, an interpolation $f$ chosen as to preserve the area $\int_{t_{i-1}}^{t_i} f(u)du =(t_i-t_{i-1}) h_i $ for a given set of values $h_i$ and knots $t_i$ is often called an \emph{histopolation}. Typically, the $h_i$ correspond to the values of a discrete histogram, often a discrete probability density. 

We can deduce that there is also an equivalence between the quadratic $C^1$ histospline representation on $h_i$ \cite[p. 79--81]{de1978practical}, and the classic cubic spline representation on the values $z_i=z_{i-1} + h_i (t_i-t_{i-1})$ at knot $t_i$. If the $(h_i)$ represent a discrete probability density, the $(z_i)$ correspond then to the discrete cumulative distribution. The equivalence holds beyond polynomial splines, as shown in \citep{bosner2014tension} for tension splines.
 

\subsection{Harmonic forwards}
The harmonic spline of \citet{fritsch1984method} can be applied directly to the forwards by replacing the forward rates $f_i$ of Equation (\ref{haganwest_forward}) with the following:
\begin{align}
\frac{1}{f_i}= \frac{t_i-t_{i-1} + 2(t_{i+1}-t_i)}{3(t_{i+1}-t_{i-1})}\frac{1}{f_{i}^d}+\frac{2(t_i-t_{i-1})+t_{i+1}-t_i}{3(t_{i+1}-t_{i-1})}\frac{1}{f_{i+1}^d}\,.
\end{align}
In \citep{lefloch2013stable}, the above was used only for $f_i^d f_{i+1}^d < 0$, otherwise $f_i$ was set to zero. With the advent of negative rates, it does not necessarily make sense to set $f_i=0$.

We may then follow the Hagan-West algorithm without explicitly enforcing any specific monotonicity and convexity constraints as, by construction, it will be already monotonic.

\subsection{Limited forwards}
\citet{huynh1993accurate} proposes a limiter approach to ensure the monotonicity of a cubic interpolation. Among the many limiters proposed, it was found in \citep{lefloch2013stable} that the rational limiter was 
 attractive. In terms of forward rates, this translates to replacing $f_i$ of Equation (\ref{haganwest_forward}) with the following:
\begin{align}
f_i= \frac{3 f_{i+1}^d f_i^d (f_{i+1}^d + f_i^d)}{(f_{i+1}^d)^2+4 f_{i+1}^d f_i^d + (f_i^d)^2}
\end{align}
Again, when the above is used for $f_i^d f_{i+1}^d < 0$ only, the forwards are guaranteed to be non-negative.

With the advent of negative rates, the Van Albada limiter may be more relevant since it guarantees positivity if the discrete forwards are positive but not if they change sign, and is a more accurate estimate. On the forward rates, it reads
\begin{align}
f_i= \frac{(f_{i+1}^d)^2 f_i^d + f_{i+1}^d (f_i^d)^2}{(f_{i+1}^d)^2+ (f_i^d)^2}\,.
\end{align}

\subsection{Lavery spline}
As a remedy to the cubic spline oscillations, \citet{lavery2000univariate} proposes to use the optimal spline under the $L^1$-norm, that is the cubic piecewise-polynomial $p$ of class $C^1$, which minimizes $\int_{t_0}^{t_n} |p''(u)| du$. In contrast, the classic $C^2$ cubic spline minimizes $\int_{t_0}^{t_n} |p''(u)|^2 du$. By discretizing the integral, the Lavery spline is the solution of a linear programming problem. While it is more involved numerically, many numerical software libraries offer fast algorithms to solve this kind of problem, for example,  GPLK or CBC.

\begin{figure}[htb]
	\begin{center}
		\includegraphics[width=0.9\textwidth]{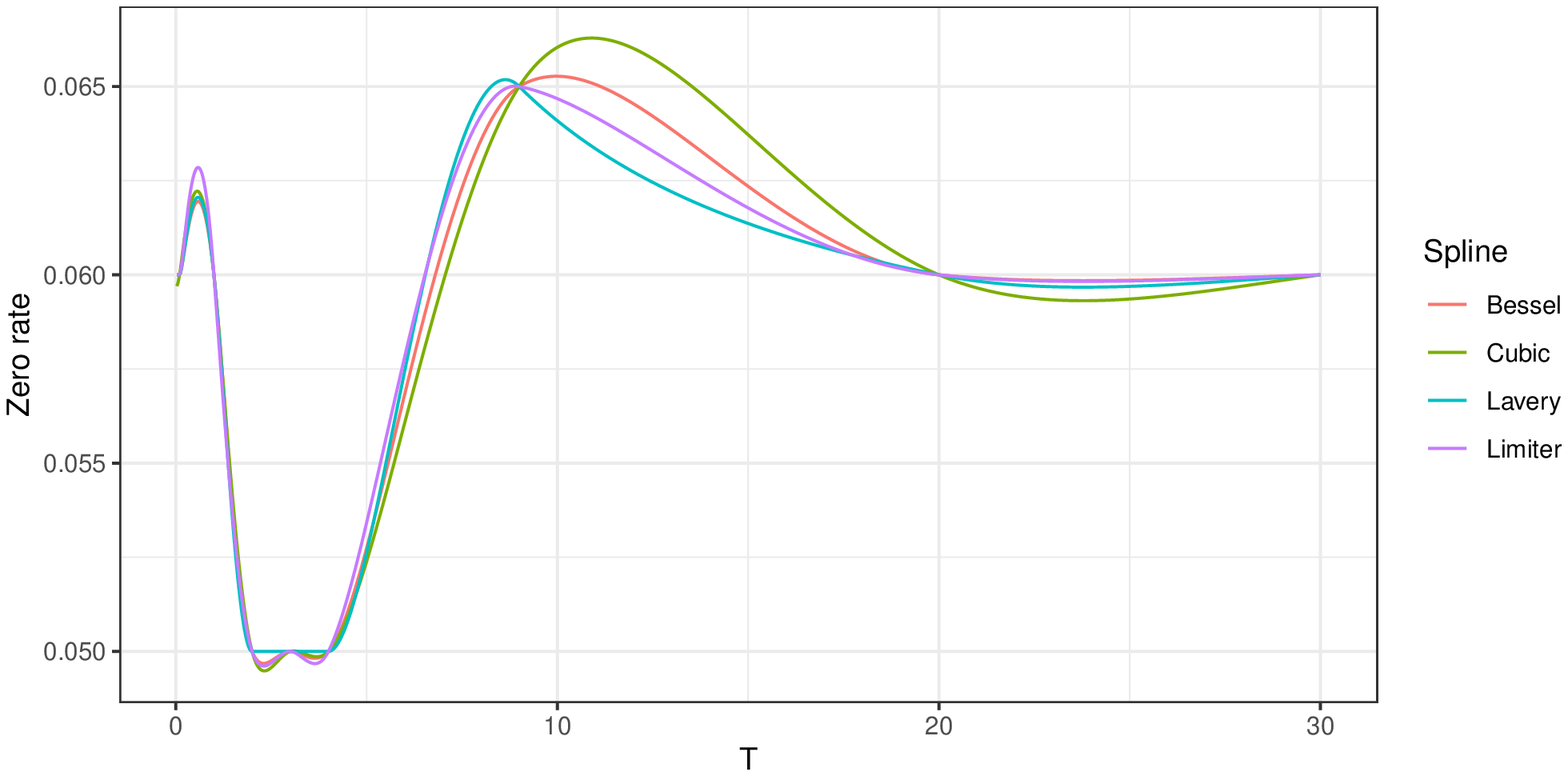}
		\caption{ \label{haganwest_zero_2006l} Zero curve for the data of \cite[Figure 4]{hagan2006interpolation}}
	\end{center}
\end{figure}

\citet{hagan2006interpolation} provide an interesting innocuous example, corresponding to their Figure 4, where a quartic spline shows major oscillations. Figure \ref{haganwest_zero_2006l} shows some undesirable small oscillations when $t \in [2, 4]$ for the Bessel, Cubic or Limited splines. 
The Lavery spline does not oscillate. While we don't display it here, the smart quadratic interpolation would lead to exactly the same curve as the Bessel spline. 
\begin{figure}[htb]
	\begin{center}
		\includegraphics[width=0.9\textwidth]{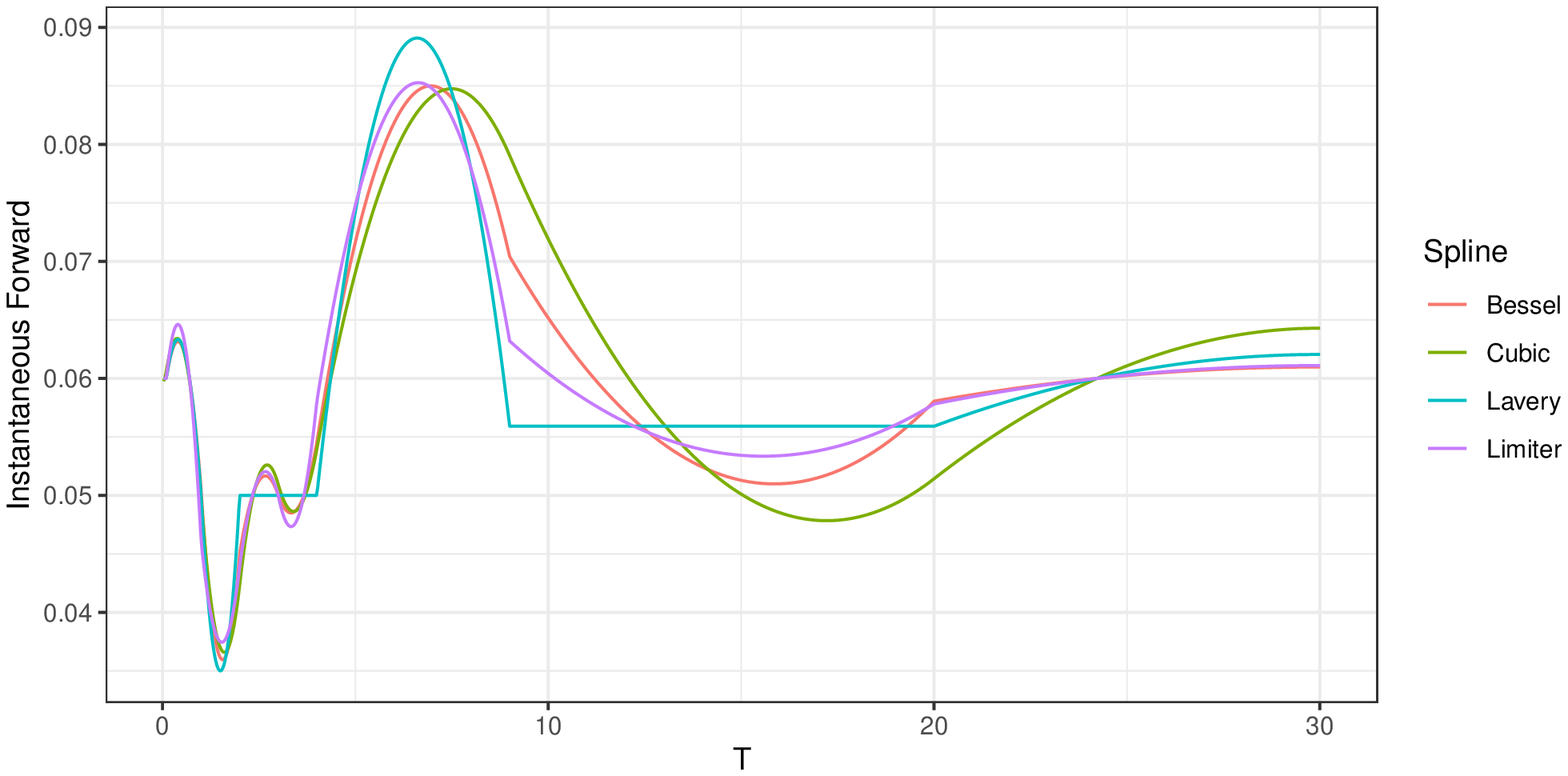}
		\caption{ \label{haganwest_forward_2006l} Forward curve for the data of \cite[Figure 4]{hagan2006interpolation}}
	\end{center}
\end{figure}

\section{Conclusion}
The smart quadratic interpolation on the forward rates corresponds exactly to the Bessel-Hermite cubic spline interpolation on the logarithm of discount factors with natural boundary conditions.

The area preserving $C^1$ quadratic interpolation on the instantaneous forward rates corresponds exactly to the $C^2$ cubic spline interpolation  on the logarithm of discount factors with natural boundary conditions.

We have also shown that a cubic spline interpolation on the logarithm of discount factors translates to a cubic spline interpolation on the discrete forward rates with a constant accrual period, but using additional knots defined in relation with the accrual period. The resulting interpolation will thus be less smooth than a direct cubic spline interpolation in terms of discrete forward rates on the forward periods start dates. This justifies the industry move towards interpolations in terms of discrete forward rates.

Finally, we note that, although the interpolation methods of \citet{hagan2006interpolation,hagan2018building} are also defined in terms of discrete forward rates, the latter must be defined on consecutive periods. As a consequence, the corresponding forward rates periods are, in most cases, not constant. The methods are thus not better adapted to the interpolation of a set of discrete forward rates of constant period, which arises in the multiple curve framework, than a classical pseudo-discount factors interpolation.

As the Hagan interpolation methods directly model the instantaneous forward rate, they become more relevant for the case of the OIS curve construction, where the one-day forward rate is very close to the instantaneous forward rate.

While the use of cubic splines is relatively standard for the yield curve construction, if the goal is to produce a very smooth yield curve, it may be more appropriate to consider smoothing splines, and particularly penalized B-splines (also known as P-splines) or alternatively, some radial basis function interpolation. We leave this for further research.
 

\funding{This research received no external funding.}
\conflictsofinterest{The authors declare no conflict of interest.}
\externalbibliography{yes}
\bibliography{forward_interpolation}
\appendixtitles{no}

\appendix
\section{Market data}
\begin{table}[h]
	\centering{
		\caption{Fed fund curve as of 2019/11/06 (data provided, courtesy of Clarus FT). \label{tbl:fedfund_curve}}
	\begin{tabular}{lrr}
		Instrument & Maturity & Par rate \\\hline
		1D OIS &	2019/11/07&	0.01560\\
		2D OIS &	2019/11/08&	0.01560\\
		Future Z19 &	2020/01/02 &	0.01560\\
		Future F20 &	2020/02/03 &	0.01535\\
		Future G20 &	2020/03/02 &	0.01495\\
		Future H20 &	2020/04/01 &	0.01475\\
		Future J20 &	2020/05/01 &	0.01440\\
		Future K20 &	2020/06/01 &	0.01425\\
		Future M20 &	2020/07/01 &	0.01405\\
		Future N20 &	2020/08/03& 	0.01385\\
		Future Q20 &	2020/09/01&	0.01370\\
		Future U20 &	2020/10/01&	0.01360\\
		OIS swap 1Y &	2020/11/09	&0.01455\\
		OIS swap 2Y &	2021/11/08	&0.01373\\
		OIS swap 3Y &	2022/11/08	&0.01354\\
		OIS swap 4Y&	2023/11/08&	0.01347\\
		OIS swap 5Y &	2024/11/08&	0.01355\\
OIS swap 6Y &	2025/11/10	&0.01375\\
		OIS swap 7Y&	2026/11/09	&0.01398\\
		OIS swap  8Y&	2027/11/08&	0.01429\\
		OIS swap 9Y	&2028/11/08	&0.01451\\
		OIS swap 10Y&	2029/11/08	&0.01484\\
		OIS swap 12Y&	2031/11/10	&0.01534\\
		OIS swap 15Y&	2034/11/08	&0.01591\\
		OIS swap 20Y&	2039/11/08	&0.01645\\
		OIS swap 25Y&	2044/11/08	&0.01662\\
		OIS swap 30Y	&2049/11/08&	0.01672\\
		OIS swap 40Y	&2059/11/10&	0.01650\\
		OIS swap 50Y	&2069/11/08	&0.01617\\\hline		
	\end{tabular}
}
\end{table}
\end{document}